\documentclass[pra,twocolumn,superscriptaddress]{revtex4-2}
\usepackage{float,graphicx,multirow, amsmath,color,dsfont}
\usepackage{amsmath,bm}
\usepackage{amssymb,mathrsfs,dsfont}
\usepackage{amssymb,mathbbol}
\usepackage{amsfonts}
\usepackage{mathrsfs}
\usepackage{float}
\usepackage{xcolor,hyperref}
\definecolor{darkblue}{rgb}{0,0.0.1,0.3}
\definecolor{darkred}{rgb}{0.6,0.1,0}
\hypersetup{colorlinks,breaklinks,
	linkcolor=darkred,urlcolor=darkblue,
	anchorcolor=darkred,citecolor=
	darkred,pdfauthor=ChSh, pdftitle=ChSh.Teleportation}
\usepackage{tikz}
\usepackage{mathbbol}
\usepackage{float}
\usepackage[normalem]{ulem}

\newcommand{\ie}{\textit{i}.\textit{e}.}

\begin{document}

	\title{Consideration of success probability and performance optimization in non-Gaussian  continuous 
		variable quantum teleportation }
	
	\author{Chandan Kumar}
	\email{chandan.quantum@gmail.com}
	\affiliation{Department of Physical Sciences,
		Indian
		Institute of Science Education and
		Research Mohali, Sector 81 SAS Nagar,
		Punjab 140306 India.}
	\author{Shikhar Arora}
	\email{shikhar.quantum@gmail.com}
	\affiliation{Department of Physical Sciences,
		Indian
		Institute of Science Education and
		Research Mohali, Sector 81 SAS Nagar,
		Punjab 140306 India.}
	\begin{abstract}
		Non-Gaussian operations have been shown to enhance the fidelity of continuous variable quantum teleportation. However, a disadvantage of these non-Gaussian operations is that they are probabilistic in nature.
		In this article, we study the trade-off between teleportation fidelity and success probability for optimal performance of the teleportation protocol, which to the best of our knowledge, has never been studied before. To this end, we first derive a unified expression for the Wigner characteristic function describing three non-Gaussian states, photon subtracted, photon added, and photon catalyzed two-mode squeezed vacuum states. We then utilize it to obtain the fidelity of teleportation for input coherent and squeezed vacuum states using the aforementioned non-Gaussian resource states. We optimize the product of the relative enhancement in fidelity and the probability of state preparation by tuning the transmissivity of the beam splitters involved in implementing non-Gaussian operations. This leads to a scenario that can be effectively implemented in a lab to enhance fidelity. It turns out that among all the considered non-Gaussian resource states, the symmetric one-photon subtracted TMSV state is the most advantageous. We provide the associated optimal squeezing and beam splitter transmissivity values for the considered non-Gaussian resource states, which will be of significant interest to the experimental community.   We also consider the effect of imperfect photon detectors on teleportation fidelity.  Further, we expect the derived Wigner characteristic function to be useful in state characterization and other quantum information processing protocols.
	\end{abstract}
	\maketitle
	\section{Introduction} 
	
	A two-mode squeezed vacuum (TMSV) state is generally employed as a resource 
	state for various continuous variable (CV) quantum information processing 
	(QIP) protocols including quantum teleportation~\cite{bk-1998}, quantum dense 
	coding~\cite{Ban_1999}, and entanglement swapping~\cite{swapping-1999}. 
	However, due to experimental limitations, it is difficult to produce high 
	squeezed states~\cite{15dB}, which puts an upper bound on the performance 
	of quantum protocols. An alternative is to perform non-Gaussian operations 
	such as photon subtraction (PS), photon addition (PA), and photon catalysis 
	(PC) on the TMSV state, which can ameliorate the nonclassicality and 
	entanglement content of the TMSV state. Further, non-Gaussian states 
	including photon-subtracted TMSV (PSTMSV), photon-added TMSV (PATMSV), and 
	photon-catalyzed TMSV (PCTMSV) states have been shown to improve the performance 
	of various QIP protocols, such as quantum  key distribution
	~\cite{qkd-pra-2013,qkd-pra-2018,qkd-pra-2019,qk2019,chandan-pra-2019,zubairy-pra-2020}, 
	quantum metrology~\cite{gerryc-pra-2012,josab-2012,braun-pra-2014,josab-2016,pra-catalysis-2021}, and quantum teleportation~\cite{tel2000,Akira-pra-2006,Anno-2007,tel2009,catalysis15,catalysis17,wang2015}. 
	This article refers to the PSTMSV, PATMSV, and PCTMSV states collectively as ``NGTMSV states" and  PS, PA, and PC operations as ``non-Gaussian operations."

	These non-Gaussian operations generated by heralding schemes using photon 
	number resolving detectors (PNRDs)~\cite{pnrd-2003,pnrd-2008} are probabilistic in nature~\cite{njp-2015}. 
	The success probability represents the fraction of successful non-Gaussian 
	operations per trial; hence, it quantifies resource utilization. When 
	considering the enhancement in quantum features such as nonclassicality, 
	entanglement, and teleportation fidelity, it is necessary to 
	account for the corresponding success probability.
	For example, the fidelity for teleporting input coherent state using PSTMSV  resource states maximizes in the unit transmissivity limit, where the success probability approaches zero [see Fig.~\ref{fpcompare}]. Therefore, focusing on maximizing fidelity renders a highly undesirable scenario for the experimental implementation of non-Gaussian quantum teleportation. To achieve optimal performance, we must trade off the enhancement in quantum features against the success probability.
	Few research studies on 
	non-Gaussian entanglement have already considered the success probability of the involved 
	non-Gaussian operations~\cite{Bartley-PRA-2013,njp-2015,Zubairy-scissor-2020,Mehdi-PRA-2020,Zubairy-22}.  Further, preparation of non-Gaussian states such as Schr\"{o}dinger cat states~\cite{Ourjoumtsev2007,quesada,Furusawa-2021} and a phase sensitivity study 
	in Mach-Zehnder interferometer using non-Gaussian states~\cite{crs-ngtmsv-met} have also taken success 
	probability into account. However, such a study in quantum 
	teleportation has not yet been undertaken. This article considers the success probability and finds an optimal balance between teleportation fidelity and success probability (resource utilization). 

	To this end, we derive the unified Wigner characteristic function of the NGTMSV states, which is then used to
	derive the analytical 
	expression for the fidelity of teleporting input coherent   and 
	squeezed vacuum states. It should be noted that calculations involving 
	non-Gaussian states are more complicated than those involving Gaussian states. 
	Additionally, the experimental scheme based on beam splitters and PNRDs
	dramatically increases the challenge for theoretical investigation as we 
	have to incorporate the free parameters associated with these devices 
	in the analytical expressions.

	We optimize the transmissivities of beam splitters to maximize the 
	teleportation fidelity. While 
	the PSTMSV and PCTMSV states can teleport input coherent and squeezed vacuum 
	states, the PATMSV states can only teleport input squeezed vacuum states with 
	large squeezing. We then study the difference between the fidelity of the 
	NGTMSV states and that of the TMSV state, $\Delta F^{\text{NG}}$, which provides us insights into the magnitudes of the 
	relative advantages provided by the different NGTMSV states.

	The maximization of teleportation fidelity renders a highly undesirable scenario in terms of success probability (resource utilization).  
	To overcome such a situation,
	we consider the success probability of non-Gaussian operations and the 
	probability of generating multi-photon Fock states; and maximize the product of $\Delta F^{\text{NG}}$ and probability of state preparation. This leads to a scenario that provides enhanced performance with optimal resource utilization. We also provide the optimal squeezing of the resource 
	states and transmissivities of the beam splitters for different NGTMSV states. 
	
	The analysis reveals that the Sym 1-PSTMSV state is the 
	most advantageous of all the considered non-Gaussian states. This can be attributed to the fact that Fock 
	states, which are generated probabilistically~\cite{mp1,mp2,mp3,mp4,mp5}, are 
	not required for PS operation. In contrast, they are required for PA and PC 
	operations, which effectively decreases the success probability of PA and PC 
	operations.    Finally, we investigate the consequences of imperfect photon detectors on teleportation fidelity. The results reveal that although the imperfect detectors lower the teleportation fidelity, it is still advantageous to implement non-Gaussian operations on the TMSV state.

	The unified  Wigner characteristic function of the NGTMSV states derived in this article is of 
	independent interest and great importance in its own right. As far as we 
	know, this expression does not exist in the literature. We expect it will help handle similar calculational challenges arising in various 
	non-Gaussian CV QIP protocols. We have explicitly provided the optimal 
	transmissivities, which shall be highly relevant to experimentalists in 
	achieving higher performance and resource optimization in non-Gaussian 
	quantum teleportation.

	The rest of the paper is organized as follows. In Sec.~\ref{sec:wig}, we 
	obtain a general expression for the Wigner characteristic function of the 
	NGTMSV states. In Sec.~\ref{sec:qt}, we provide a comprehensive study of the 
	teleportation of input coherent and squeezed vacuum states.   In Sec.~\ref{sec:imp}, we analyze the
		effect of imperfect detectors on teleportation fidelity.  Finally, 
	we conclude with a discussion in Sec.~\ref{sec:conc}, where we outline the 
	implications and future aspects of the current work. In the Appendix, we 
	briefly describe the phase space description of the CV systems relevant to 
	this article.

	
	\section{Wigner characteristic function of the NGTMSV states}\label{sec:wig}

	\begin{figure}[h!]
		\includegraphics[scale=1]{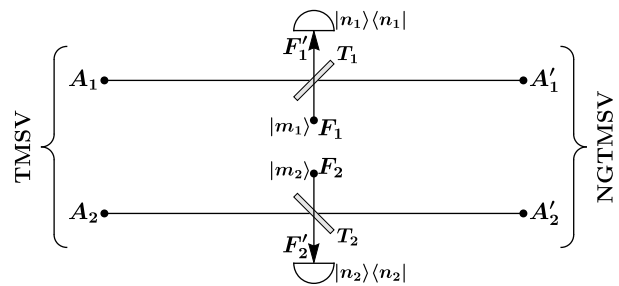}
		\caption{Experimental setup for the preparation of non-Gaussian TMSV 
			state.  The TMSV state is mixed with ancilla modes initiated to 
			Fock states $|m_1\rangle$ and $|m_2\rangle$ using beam splitters. 
			Simultaneous detections of $n_1$ and $n_2$ photons in the output 
			modes corresponding to the ancilla modes herald successful 
			non-Gaussian operations on both modes.}
		\label{figsub}
	\end{figure}

	The experimental setup for the generation of NGTMSV states is shown in 
	Fig.~\ref{figsub}. We consider a TMSV state labeled by $A_1$ and $A_2$. 
	We represent the modes $A_1$ and $A_2$ by the quadrature operators 
	$(\hat{q}_1,\hat{p}_1)^T$ and  $(\hat{q}_2,\hat{p}_2)^T$, respectively. 
	  These quadrature operators are related to the annihilation and creation operators of the $i^{\text{th}}$ through the relation 
		$\hat{a}_i= (\hat{q}_i+i\hat{p}_i)/\sqrt{2}$ and $ \hat{a}^{\dagger}_i=  (\hat{q}_i-i\hat{p}_i)/\sqrt{2}$.
		The TMSV state is obtained by
		the action of two-mode squeezing operator on two single-mode vacuum state~\cite{tmsv1,tmsv2,Reid}:
		\begin{equation}
			|\psi\rangle_{A_1A_2}=	\exp[r (\hat{a}_1^{\dagger} \hat{a}_2^{\dagger}-\hat{a}_1
			\hat{a}_2) ]|0\rangle_1|0\rangle_2,
		\end{equation}
		where $r$ is the squeezing parameter.  
	We also consider two auxiliary modes labeled by $F_1$ and $F_2$ and 
	initiated to Fock states $|m_1\rangle$ and $|m_2\rangle$, respectively.  
	We represent the modes $F_1$ and $F_2$ by the quadrature operators 
	$(\hat{q}_3,\hat{p}_3)^T$ and  $(\hat{q}_4,\hat{p}_4)^T$, respectively. 
	Mode $A_1$ ($A_2$) is mixed with mode $F_1$ ($F_2$) using beam-splitter 
	of transmissivity $T_1$  ($T_2$).    For convenience in calculating teleportation fidelity, we work in the phase space formalism using Wigner characteristic function.
		The Wigner 
		characteristic function for a density operator 
		$\hat{\rho}$ of an $n$-mode quantum system  can be calculated as follows:
		\begin{equation}\label{wigde}
			\chi(\Lambda) = \text{Tr}[\hat{\rho} \, \exp(-i \Lambda^T \Omega \hat{\xi})],
		\end{equation}
		where $\xi = (\hat{q_1}, \hat{p_1},\dots \hat{q_n}, \hat{p_n})^T$,  
		$\Lambda = (\Lambda_1, \Lambda_2, \dots \Lambda_n)^T$ with  
		$\Lambda_i = (\tau_i, \sigma_i)^T \in \mathcal{R}^2$ and  $\Omega$
			is the symplectic form on $n$-mode quantum system.  
	Prior to the beam splitter operations, 
	the Wigner characteristic function of the four-mode system is given by 
	\begin{equation}
		\chi_{F_1 A_1 A_2 F_2}(\Lambda) =  \chi_{A_1 A_2}(\Lambda) \chi_{|m_1\rangle}(\Lambda_3)  \chi_{|m_2\rangle}(\Lambda_4),
	\end{equation}
	where $\chi_{|m_1\rangle}(\Lambda_3)$ is the Wigner characteristic function of 
	the Fock state $|m_1\rangle$ and $\chi_{|m_2\rangle}(\Lambda_4)$ is the 
	Wigner characteristic function of the Fock state $|m_2\rangle$. The four modes 
	get entangled as a result of the mixing of the modes by the two beam splitters 
	collectively represented by the symplectic transformation matrix 
	$ B(T_1,T_2)=B_{A_1 F_1}(T_1) \, {\oplus} \, B_{A_2 F_2} (T_2)$, where $B_{ij}(T)$ is 
	the beam splitter operation given in Eq.~(\ref{beamsplitter}) of the 
	Appendix~\ref{intro}. The evolved Wigner characteristic function can be 
	readily evaluated using Eq.~(\ref{transformation}) of the Appendix~\ref{intro}:
	\begin{equation}
		\chi_{F_1' A_1' A_2' F_2'}(\Lambda)  =\chi_{F_1 A_1 A_2 F_2}( B(T_1,T_2)^{-1}\Lambda).
	\end{equation}
	PNRDs, represented by the positive-operator-valued measure (POVM) 
	$\{\Pi_{n_1}=|n_1\rangle\langle n_1|,\mathbb{1}-\Pi_{n_1}\}$, and 
	$\{\Pi_{n_2}=|n_2\rangle\langle n_2|,\mathbb{1}-\Pi_{n_2}\}$, are used to 
	measure the transformed auxiliary modes $F_1^{'}$ and $F_2^{'}$,  respectively. 
	The simultaneous click of the POVM elements $\Pi_{n_1}$ and $\Pi_{n_2}$ heralds 
	successful non-Gaussian operations on both modes. The post-measurement 
	state corresponds to the unnormalized Wigner characteristic function of the 
	NGTMSV states:
	\begin{equation}\label{detect}
		\begin{aligned}
			\widetilde{\chi}^{\text{NG}}_{A_1' A_2'}=& \frac{1}{(2 \pi)^2} \int d^2 \Lambda_3  d^2 \Lambda_4 
			\underbrace{\chi_{F_1' A_1' A_2' F_2'}(\Lambda)}_{\text{Four mode entangled state}}\\
			&\times 
			\underbrace{\chi_{|n_1\rangle }(\Lambda_3)}_{\text{Projection on }|n_1\rangle \langle n_1|} 
			\underbrace{\chi_{|n_2\rangle }(\Lambda_4)}_{\text{Projection on }|n_2\rangle \langle n_2|}. \\
		\end{aligned}
	\end{equation}
	We can appropriately choose input and measured photons in the 
	auxiliary modes and perform different non-Gaussian operations. For instance, 
	by choosing $m_i {<} n_i$, $m_i {>} n_i$, and $m_i {=} n_i$, we can perform PS, PA, 
	and PC operations on mode $A_i$, respectively. These operations on the TMSV state 
	result in the generation of non-Gaussian states, abbreviated 
	as PSTMSV, PATMSV, and PCTMSV states. In our analysis, we consider $m_1=m_2=0$ 
	and $n_1=n_2=0$ for PS and PA operations, respectively. Further, we can 
	perform asymmetric and symmetric non-Gaussian operations on the TMSV state as 
	illustrated in Table~\ref{table1}. Here we note that we perform the 
	asymmetric non-Gaussian operations on mode $A_2$ of the TMSV state, 
	and therefore, only beam splitter transmissivity $T_2$ appears in 
	relevant expressions.

	\begin{table}[h!]
		\centering
		\caption{\label{table1} 
			Restrictions on the input photons ($m_i$) and the number of photons 
			measured ($n_i$) in the auxiliary modes for various asymmetric and 
			symmetric non-Gaussian operations on the TMSV state.}
		\setlength{\tabcolsep}{7pt}
		\renewcommand{\arraystretch}{1}	
		\begin{tabular}{ |c|c c|c c|}
			\hline 
			\multirow{2}{*}{Operations} & \multicolumn{2}{c|}{Input} &  \multicolumn{2}{c|}{Detected}\\
			& $m_1$ & $m_2$ & $n_1$ & $n_2$ \\
			\hline \hline
			Asym $n$-PS  & 0& 0& 0&$n$ \\ \cline{1-5}
			Asym $n$-PA  & 0 & $n$ & 0 & 0 \\ \cline{1-5}
			Asym $n$-PC  & 0 & $n$ & 0&$n$  \\ \hline  \hline
			Sym $n$-PS 	 & 0& 0& $n$ &$n$ \\ \cline{1-5}
			Sym $n$-PA 	 & $n$ & $n$ & 0 & 0 \\ \cline{1-5}
			Sym $n$-PC 	 & $n$ & $n$ & $n$ & $n$ \\  \hline \hline
		\end{tabular}
	\end{table}

	Using the Wigner characteristic function of the Fock state~(\ref{charfock1}) 
	and  integrating Eq.~(\ref{detect}), we get 
	\begin{equation}\label{eqchar}
		\begin{aligned}
			\widetilde{\chi}^{\text{NG}}_{A_1' A_2'}&= \bm{\widehat{F}_1} \exp 
			\left(\bm{\Lambda}^T M_1 \bm{\Lambda}+\bm{u}^T M_2 \bm{\Lambda} + \bm{u}^T M_3 \bm{u} \right),
		\end{aligned}
	\end{equation}
	where the column vectors $\bm{\Lambda}$ and $\bm{u}$ are defined as  
	$(\tau_1,\sigma_1,\tau_2,\sigma_2)^T$ and $(u_1,v_1,u_2,v_2,u_1',v_1',u_2',v_2')^T$ 
	respectively, and the matrices $M_1$, $M_2$ and $M_3$ are given in 
	Eqs.~(\ref{m1}),~(\ref{m2}) and~(\ref{m3}) of the Appendix~\ref{appex}. 
	Further,  the differential operator $\bm{\widehat{F}_1} $ is defined as 
	\begin{equation}
		\begin{aligned}
			\bm{\widehat{F}_1} = \frac{2^{-(m_1+m_2+n_1+n_2)}}{m_1!m_2!n_1!n_2!} \frac{\partial^{m_1}}{\partial\,u_1^{m_1}} \frac{\partial^{m_1}}{\partial\,v_1^{m_1}} \frac{\partial^{m_2}}{\partial\,u_2^{m_2}} \frac{\partial^{m_2}}{\partial\,v_2^{m_2}}\\
			\times \frac{\partial^{n_1}}{\partial\,u_1'^{n_1}} \frac{\partial^{n_1}}{\partial\,v_1'^{n_1}} \frac{\partial^{n_2}}{\partial\,u_2'^{n_2}} \frac{\partial^{n_2}}{\partial\,v_2'^{n_2}} \{ \bullet \}_{\substack{u_1= v_1=u_2= v_2=0\\ u_1'= v_1'=u_2'= v_2'=0}}.\\
		\end{aligned}
	\end{equation}
	The normalization factor corresponding to Eq.~(\ref{eqchar}) represents 
	the probability of success of non-Gaussian operations in both the modes 
	and is evaluated as
	\begin{equation}\label{probng}
		\begin{aligned}
			P^{\text{NG}}&=\widetilde{\chi}^{\text{NG}}_{A_1' A_2'}\bigg|_{\tau_1= \sigma_1= \tau_2= \sigma_2=0}
			= \bm{\widehat{F}_1} \exp 
			\left(\bm{u}^T M_3 \bm{u} \right). \\		
		\end{aligned}
	\end{equation}
	The normalized  Wigner characteristic function $\chi^{\text{NG}}_{A'_1 A'_2}$ 
	of NGTMSV states is obtained as
	\begin{equation}\label{normPS}
		\chi^{\text{NG}}_{A'_1 A'_2}(\tau_1,\sigma_1,\tau_2,\sigma_2) ={\left(P^{\text{NG}}\right)}^{-1}\widetilde{\chi}^{\text{NG}}_{A_1' A_2'}(\tau_1,\sigma_1,\tau_2,\sigma_2).
	\end{equation}
	Wigner characteristic function of several special states can be obtained from 
	Eq.~(\ref{normPS}) as limiting cases. By taking the limit  $T_1 \rightarrow 1$ and $T_2 \rightarrow 1$ in the symmetric 
	PS case with $m_1=m_2=0$, we obtain  the Wigner characteristic function of the 
	ideal PSTMSV state $\hat{a}_1^{n_1} \hat{a}_2^{n_2}  |\text{TMSV}\rangle$. 
	Similarly, taking the limit $T_1 \rightarrow 1$ and $T_2 \rightarrow 1$ in the symmetric PA case   with $n_1=n_2=0$ renders 
	the Wigner characteristic function of the ideal PATMSV state 
	$\hat{a}{_1^{\dagger }}^{m_1}\hat{a}{_2^{\dagger }}^{m_2}  |\text{TMSV}\rangle$.

	
	\section{Teleportation  using NGTMSV resource states}\label{sec:qt}

	Having derived the Wigner characteristic function of the  NGTMSV states, we 
	proceed to derive the fidelity for teleporting input coherent and squeezed 
	vacuum states. We follow the Braunstein-Kimble (BK) protocol for teleporting 
	an unknown input quantum state between two distant physical 
	systems~\cite{bk-1998}. To begin with, an entangled resource is shared 
	between Alice and Bob. An unknown input quantum state to be teleported is 
	provided to Alice. The density operator of the entangled resource state 
	and the unknown input state is represented by $\rho_{A_1' A_2'}$ and  
	$\rho_{\text{in}}$, respectively. Their representation in terms of Wigner 
	characteristic function are $\chi_{A_1' A_2'}(\Lambda_1, \Lambda_2)$ and 
	$\chi_{\text{in}}(\Lambda_{\text{in}})$, respectively.

	Alice combines her mode and the single-mode input state using a balanced 
	beam splitter. After that, the two output modes of the beam splitter are 
	subjected to homodyne measurement by Alice, and the results are classically 
	communicated to Bob. Based on the results, Bob displaces his mode $A_2'$, 
	and the resultant mode is denoted by `out'. The mode `out'  corresponds to 
	the teleported state. The Wigner characteristic function allows us to write 
	the teleported state as a product of the input state and the entangled 
	resource state~\cite{Marian-pra-2006}:
	\begin{equation}\label{teleported}
		\chi_{\text{out}}(\tau_2,\sigma_2) = \chi_{\text{in}}(\tau_2,\sigma_2) \chi_{A_1' A_2'}(\tau_2,-\sigma_2,\tau_2,\sigma_2).
	\end{equation}
	We define fidelity of teleportation as the overlap between the single mode  
	input state $\rho_{\text{in}}$ and the teleported state $\rho_{\text{out}}$ 
	to quantify the success of the  protocol:
	\begin{equation}\label{fidex1}
		\begin{aligned}
			F &=\text{Tr} [\rho_{\text{in}}\rho_{\text{out}}],\\
			&=\frac{1}{2 \pi} \int d^2 \Lambda_2  \chi_{\text{in}}(\Lambda_2)
			\chi_{\text{out}}(-\Lambda_2).
		\end{aligned}
	\end{equation}
	It has been shown that a maximum fidelity of $1/2$ can be achieved without 
	using a shared entangled state for teleporting input coherent state~\cite{Braunstein-jmo-2000,Braunstein-pra-2001}. 
	Hence, successful quantum teleportation is marked by the magnitude of the 
	fidelity rising above the classical limit of $1/2$. We require an infinitely 
	entangled resource state to achieve perfect teleportation with unit fidelity.
	   Further, the fidelity should be greater than 2/3 for secure teleportation, \ie, for ensuring that any copy of  the teleported state is not available with an eavesdropper~\cite{Cerf, Grangier,Paris-pra-2003}. 
	
	\subsection{Teleporting an input coherent state} 
	We now move on to compute the fidelity for teleporting an input coherent 
	state via NGTMSV resource states~(\ref{normPS}). Using the Wigner 
	characteristic function of the coherent state [Eq.~(\ref{chi_coh}) of 
	Appendix~\ref{intro}], the fidelity can be evaluated using Eq.~(\ref{fidex1}), 
	which turns out to be
	\begin{equation}\label{ngfidc}
		\begin{aligned}
			&F^\text{NG}_\text{coh}= \bm{\widehat{F}_1} \exp 
			\left(\bm{u}^T M_4 \bm{u} \right),
		\end{aligned}
	\end{equation}
	where the matrix $M_4$ is given in Eq.~(\ref{m4}) of the Appendix~\ref{appex}.
	By  taking the limit  $T_1 \rightarrow 1$ and $T_2 \rightarrow 1$ with $m_1=n_1$, $m_2=n_2$  in Eq.~(\ref{ngfidc}), we obtain 
	the fidelity of quantum teleportation using the TMSV resource state:
	\begin{equation}\label{fidtmsc}
		F^\text{TMSV}=\frac{1+\tanh{r}}{2}.
	\end{equation}
	\begin{figure}[h!] 
		\begin{center}
			\includegraphics[scale=1]{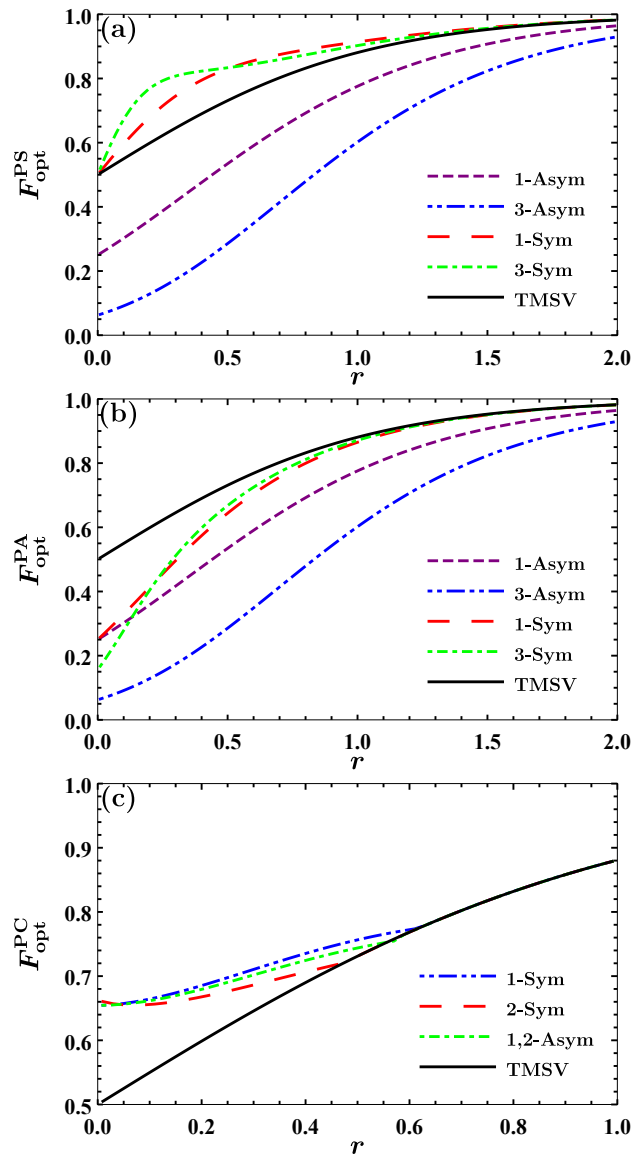}
			\caption{Optimized fidelity for teleporting input coherent state as a 
				function of the squeezing parameter for (a) PSTMSV states, 
				(b) PATMSV states, and
				(c) PCTMSV states.  The transmissivities of the
				beam splitters are optimized in order to maximize the fidelity.  }
			\label{fidc}
		\end{center}
	\end{figure}
	
	After deriving the analytical expression of fidelity for teleporting input 
	coherent state, we now proceed to numerical investigation of the fidelity. 
	We first numerically optimize the transmissivities of the beam splitters to 
	maximize the fidelity, and the results are shown in Fig.~\ref{fidc}. 
	
	While symmetric PSTMSV states show an advantage over the TMSV 
	state, asymmetric PSTMSV states underperform as compared 
	to the TMSV state. Further, the optimal transmissivity is one in this case; 
	hence, the results correspond to ideal PSTMSV states.  
	
	We then observe that neither symmetric nor asymmetric PATMSV states
	improve the performance compared to the TMSV state. In this case, the 
	optimal transmissivity also turns out to be one; therefore, these results 
	correspond to ideal PATMSV states. We note that the quantum 
	teleportation using ideal PSTMSV and PATMSV states as a resource has 
	already been investigated in Refs.~\cite{Anno-2007,wang2015}.
	
	Finally, we observe that only symmetric PCTMSV states 
	enhance fidelity over the TMSV state. As can be seen in 
	Fig.~\ref{fidc}(c), this enhancement is observed till a certain squeezing 
	threshold beyond which the fidelity is optimized at unit transmissivity. 
	As we can see from the schematic in Fig.~\ref{figsub} that the output 
	state at unit transmissivity in the case of PC operation is simply the 
	TMSV state. Therefore, the fidelity at unit transmissivity is equal to 
	that of the TMSV state. This optimization of fidelity for PCTMSV states 
	has been performed in Ref.~\cite{catalysis17}; however, their results for 
	PCTMSV states show a lower fidelity than the TMSV state in the region 
	where our results show equal fidelity for PCTMSV and TMSV states. 
	Further, the Asym 1,2-PCTMSV state also enhances the fidelity compared 
	to the TMSV state. We note that Asym 1,2-PCTMSV state can be generated by 
	setting the parameters as $m_1=n_1=1$ and $m_2=n_2=2$.

	\subsubsection{Relative enhancement in fidelity}
	In the previous section, we studied the absolute performance of the NGTMSV states. 
	  Let us consider, for instance, the case of PSTMSV resource states [Fig. 2(a)], where the fidelity is maximized for high values of squeezing. For large values of squeezing, the original TMSV state performs almost the same as the PSTMSV state. Therefore, performing the non-Gaussian operation in the high squeezing range is not preferable.
		In order to find the squeezing value where implementing non-Gaussian operation provides the maximum advantage,
		we define a figure of merit as the difference in teleportation fidelity between the NGTMSV states and the TMSV 
		state as
		\begin{equation}
			\Delta F^{\text{NG}} = F^{\text{NG}}-F^{\text{TMSV}} .
		\end{equation}
		The squeezing parameter that renders maximum $\Delta F^{\text{NG}}$ corresponds to the maximum advantage in implementing non-Gaussian operations. 
	We optimize $\Delta F^{\text{NG}}$ over the transmissivity parameters and 
	since $F^{\text{TMSV}}$ is independent of transmissivity, the optimized 
	value of $\Delta F^{\text{NG}}$ can be simply given by
	\begin{equation}\label{dif-fidelity}
		\Delta F_{\text{opt}}^{\text{NG}} = F^{\text{NG}}_{\text{opt}}-F^{\text{TMSV}} .
	\end{equation}
	\begin{figure}[h!] 
		\begin{center}
			\includegraphics[scale=1]{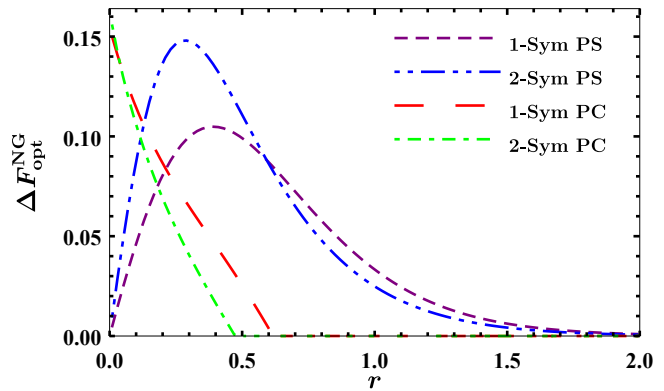}
			\caption{The optimized difference of teleportation fidelity between the
				NGTMSV  and the TMSV  states, $\Delta F_{\text{opt}}^{\text{NG}}$,  
				as a function of the squeezing parameter for different non-Gaussian 
				states. The transmissivities have been optimized to maximize 
				$\Delta F^{\text{NG}}$.  }
			\label{dfidc}
		\end{center}
	\end{figure}
	We plot $\Delta F_{\text{opt}}^{\text{NG}}$ as a function of squeezing 
	parameter in Fig.~\ref{dfidc} for the symmetric PSTMSV and 
	PCTMSV states. While $\Delta F_{\text{opt}}$ for PSTMSV 
	states is maximized at an intermediate squeezing, $\Delta F_{\text{opt}}$ for 
	PCTMSV states is maximized in the limit of zero squeezing. 
	We have not considered PATMSV 
	states as they do not provide any advantage over the original TMSV state. 
	We shall see later in this article that PATMSV states may be advantageous 
	over the TMSV state in the teleportation of an input squeezed vacuum state.

	\begin{figure}[h!]
		\begin{center}
			\includegraphics[scale=1]{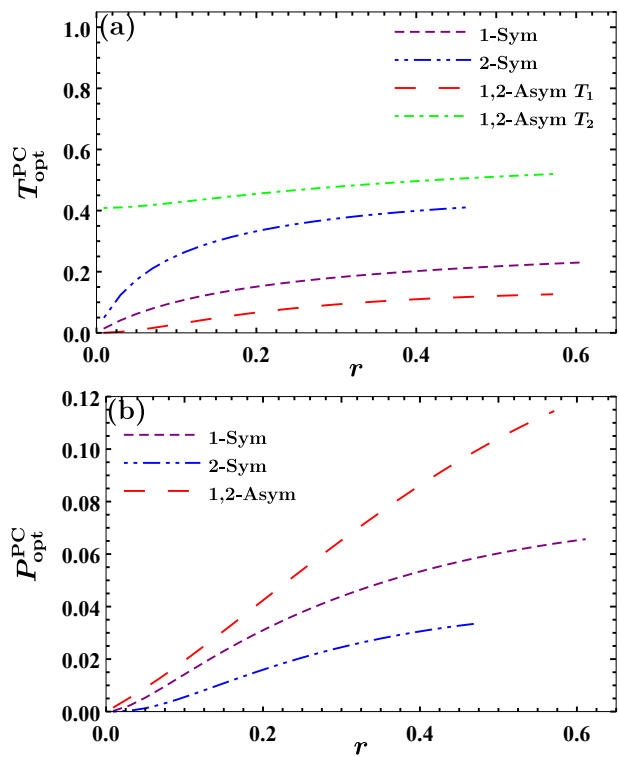}
			\caption{(a) Optimal beam splitter transmissivities and (b) corresponding probabilities as a function 
				of the squeezing parameter for different PCTMSV states.
				The beam splitter transmissivities and corresponding probabilities have been truncated at the 
				minimum squeezing where the TMSV state and the PCTMSV state have 
				the same fidelity.  }
			\label{taufidc}
		\end{center}
	\end{figure}

	As evident from Eq.~(\ref{dif-fidelity}), the optimal 
	transmissivities of the beam splitters maximizing $ F^{\text{NG}}$ and $\Delta F^{\text{NG}}$ are same.
	As mentioned earlier, the fidelity is maximized in the unit transmissivity limit for the PSTMSV and PATMSV states, irrespective of the squeezing. For the PCTMSV states, the optimal transmissivities as a function of the squeezing parameter are 
	shown in Fig.~\ref{taufidc}(a). While the fidelity for the Sym $n$-PCTMSV states is 
	maximized for $T_1 =T_2$, the fidelity for the   Asym 1,2-PCTMSV state is maximized when $T_1 \neq T_2$.

	The success probability of non-Gaussian operations, $P^{\text{NG}} $, which can be thought of as the fraction of 
	successful non-Gaussian operations per trial, quantifies the resource 
	utilization. For PSTMSV and PATMSV states, the success probability approaches zero in the unit transmissivity limit. 
	Here the resource utilization approaches zero; therefore, this scenario is highly undesirable from an experimental point of view. Further, for the PCTMSV states, we have shown the success probability corresponding to the optimal transmissivity as a function of the squeezing parameter in Fig.~\ref{taufidc}(b). As we observed in Fig.~\ref{dfidc}, the maximum enhancement in fidelity is obtained in the zero squeezing limit for the PCTMSV resource states, but the corresponding success probability approaches zero. Therefore, this situation is also undesirable.

	\subsubsection{Relative enhancement in fidelity per trial}

\begin{figure}[h!]
		\begin{center}
			\includegraphics[scale=1]{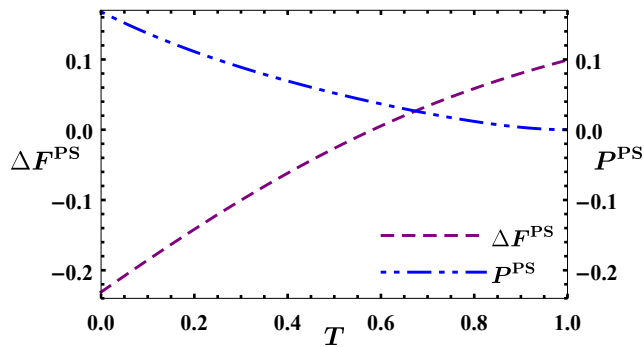}
			\caption{The fidelity difference $\Delta F^{\text{PS}}$ between the
				Sym 1-PSTMSV and the TMSV  states, and 
				preparation probability $P^\text{PS}$ of the Sym 1-PSTMSV state as a function of transmissivity $T$ of the beam splitter. The squeezing of the resource state is taken to be $r=0.5$.}
			\label{fpcompare}
		\end{center}
	\end{figure}

	As we saw in the previous section, working at optimal squeezing and transmissivity 
	parameters maximizing $\Delta F^{\text{NG}}$ may not represent the best scenario. 
	To visualize this explicitly, we show $\Delta F^{\text{PS}}$ and $P^{\text{PS}}$ as a function 
	of the transmissivity for the Sym 1-PSTMSV 
	resource state in Fig.~\ref{fpcompare}. While $\Delta F^{\text{PS}}$ is maximized in the unit transmissivity limit, the success probability approaches zero,  which is an undesirable scenario for the generation of non-Gaussian state.
	To find an optimal scenario, we can trade-off between the success probability and $\Delta F^{\text{PS}}$  by adjusting the transmissivity.

	Similarly, to find the optimal scenario for different NGTMSV states, we can 
	trade-off between the success probability and $\Delta F^{\text{NG}}$  
	by adjusting the transmissivities of the beam splitters for a given squeezing.
	In the following analysis, we find optimal transmissivity parameters that 
	renders the product of $\Delta F^{\text{NG}}$ and $P^{\text{NG}}$ maximum.

	\begin{figure}[h!]
		\begin{center}
			\includegraphics[scale=1]{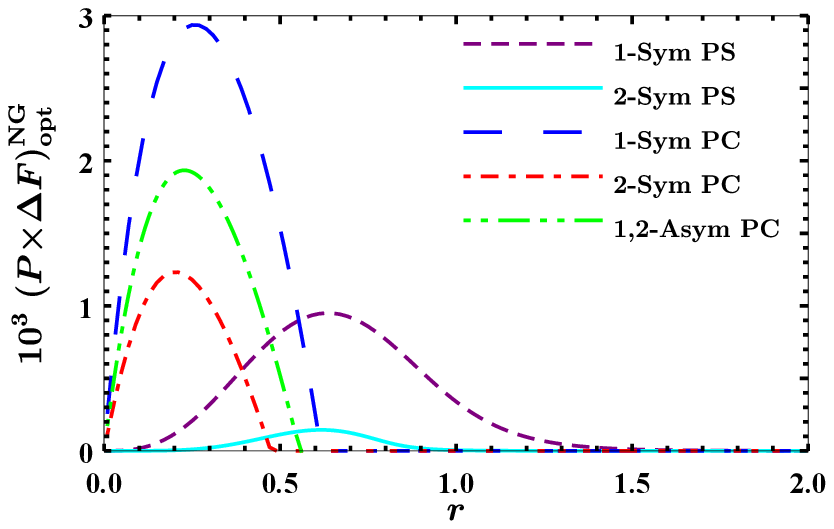}
			\caption{   The optimized product 
				$(P {\times} \Delta F)^{\text{NG}}$ as a function 
				of the squeezing parameter. The transmissivities have been 
				optimized to maximize the product. }
			\label{pdfidc}
		\end{center}
	\end{figure}
	
	We plot the optimized product 
	$(P {\times} \Delta F)_{\text{opt}}^{\text{NG}}$  as a function 
	of squeezing in Fig.~\ref{pdfidc}. The results reveal that 1-Sym 
	PC operation provides the maximum advantage when the success probability is 
	taken into account. Further, among the considered non-Gaussian operations in 
	Fig.~\ref{pdfidc}, all different PC operations outperform the PS 
	operations.

\begin{figure}[h!] 
		\begin{center}
			\includegraphics[scale=1]{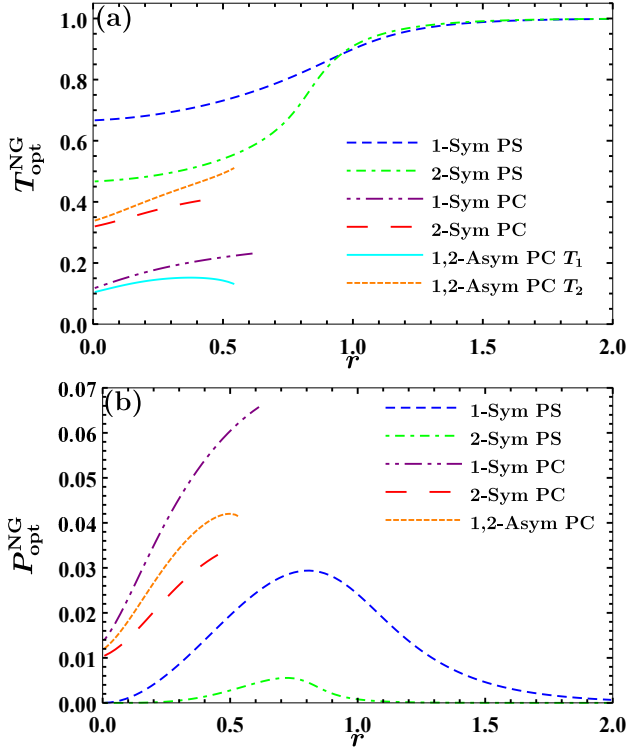}
			\caption{(a) Optimal beam splitter transmissivities and (b) corresponding probabilities maximizing the 
				product   $P^{\text{NG}} {\times} \Delta F^{\text{NG}}$ as a function
				of the squeezing parameter for different NGTMSV
				states. The beam splitter transmissivities and corresponding probabilities have been
				truncated at the minimum squeezing where the TMSV state and
				the PCTMSV states have the same fidelity.}
			\label{taupdfidc}
		\end{center}
	\end{figure}

	Figure~\ref{taupdfidc} shows  the optimal beam splitter 
	transmissivities and the corresponding success probabilities, maximizing the product  
	$P^{\text{NG}} {\times} \Delta F^{\text{NG}}$ as a function of the squeezing parameter.
	We notice that the success probability of the Sym 1-PSTMSV state is $\approx 3 \%$    at the squeezing parameter maximizing the product of $\Delta F^{\text{NG}}$ and $P^{\text{NG}}$. Similarly, the success probability of the Sym 1-PCTMSV state is $\approx 4 \%$. These success probabilities are significantly larger than those obtained while maximizing $\Delta F^{\text{NG}}$. By working at optimal conditions, these non-Gaussian operations can be effectively implemented in a lab for fidelity enhancement.

	  We compare our results with previous investigations on entanglement which considered the optimization of  $\Delta E^{\text{NG}} \times P^{\text{NG}}$, where `E' stands for entanglement~\cite{Bartley-PRA-2013}. It was shown that  1-Asym PC operation provides maximum advantages for low squeezing while 1-Asym PA operation provides maximum advantages for high squeezing.

	As shown in Fig.~\ref{figsub}, the auxiliary modes are initialized to Fock 
	states in PA and PC operations, in contrast to vacuum states ($|00\rangle_{A_1A_2}$)   in PS operation. These Fock states can be generated by 
	photon-number measurement on one mode of the TMSV state. The success 
	probability of producing Fock state $|m\rangle$ is
	\begin{equation}
		P_{|m\rangle} = (1-\lambda^2)\lambda^{2m}, \quad \text{with} \quad \lambda = \tanh r.
	\end{equation}
	Therefore, the effective probability is defined as
	\begin{equation}
		P_{\text{eff}} =(P_{|m_1\rangle} P_{|m_2\rangle} )P^{\text{NG}}.
	\end{equation}

	\begin{figure}[h!] 
		\begin{center}
			\includegraphics[scale=1]{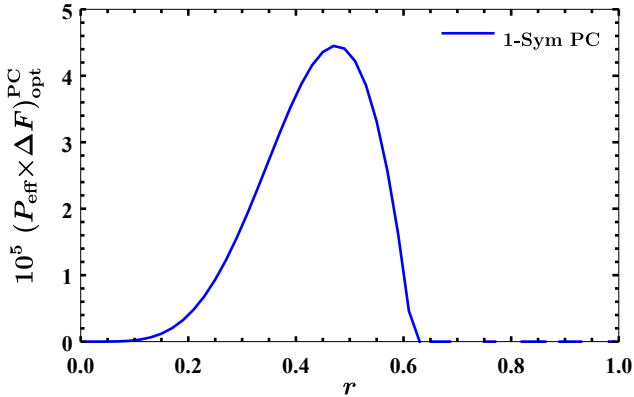}
			\caption{ The optimized product 
				$(P_{\text{eff}} {\times} \Delta F)_{\text{opt}}^{\text{PC}}$ 
				as a function of the squeezing parameter. The transmissivities have 
				been optimized to maximize the product.}
			\label{ppdfidc}
		\end{center}
	\end{figure}
	
		   Parameters maximizing the product $\Delta F^{\text{NG}} \times P_{\text{eff}}$ enable us to  assess the efficacy of the different non-Gaussian resource states.  To obtain the optimal parameters, we  optimize the product $P_{\text{eff}}^{\text{PC}} {\times} \Delta F^{\text{PC}}$ 
	over the transmissivity parameters and the result is shown in Fig.~\ref{ppdfidc}.
	For 1-Sym PC operation, the maximum value of the optimized product 
	$P_{\text{eff}} {\times} \Delta F_{\text{opt}}$ is $\approx 5 {\times} 10^{-5}$. 
	In contrast, for 1-Sym PS operation, the maximum value of the optimized product 
	$P {\times} \Delta F_{\text{opt}}$ is $\approx 1{\times} 10^{-3}$. Therefore, 
	1-Sym PS operation (Sym 1-PSTMSV state) turns out to be the most beneficial 
	operation (state) for teleporting an input coherent state.    The major factor behind this result is the fact that
		 multiphoton Fock states $|m_1 m_2\rangle_{A_1A_2}$ are required for the PC and PA operation, whereas only vacuum states $|00\rangle_{A_1A_2}$  are required for the PS operation. When we take the probability $P_{|m_1\rangle} P_{|m_2\rangle}$ to generate Fock states  into account, the effective probability  $  P_{\text{eff}} =  (P_{|m_1\rangle} P_{|m_2\rangle} )P^{\text{NG}}$ for PA  and PC operations becomes very low as compared to PS operation. To see this explicitly, we have added Table~\ref{table2} , showing the   magnitudes of $\Delta F^{\text{NG}} $ and $P_{\text{eff}}$ at optimal parameters [Figs.~\ref{pdfidc} and \ref{ppdfidc}].
		
		While the fidelity is of the same order for 1-Sym PS and 1-Sym PC operations~\cite{wang2015,catalysis17},  the magnitude of $P_{\text{eff}} =(P_{|1\rangle} P_{|1\rangle} )P^{\text{PC}} $   (for PC operation) is one order low as compared to   that  of  $P^{\text{PS}}$  and $P^{\text{PC}}$.
		 This renders the PS operation more advantageous compared to PC operation.

	   Upon accounting for the probability of Fock state generation $P_{|m_1\rangle} P_{|m_2\rangle}$, the entanglement study  of Ref.~\cite{Bartley-PRA-2013}  observes  that 1-Asym PC operation remains the maximum advantageous operation for low squeezing; however, for high squeezing, 1-Asym PS operation provides  the  maximum advantage. 
	
	\begin{table}[h!]
		\centering
		\caption{\label{table2}
		  Maximum value of the product 	$(P {\times} \Delta F)_{\text{opt}}^{\text{NG}}$ and corresponding  $P$ and $\Delta F$. }
		\renewcommand{\arraystretch}{1.5}
		\begin{tabular}{ |c |c |c|c|}
			\hline \hline
			1-Sym PS & $(P {\times} \Delta F )_{\text{max}}\approx 1{\times} 10^{-3}$ 
			  & $P \approx 0.03$ & $\Delta F \approx 0.04$ \\ \hline
			   1-Sym PC & $(P {\times} \Delta F )_{\text{max}}\approx 3{\times} 10^{-3}$ 
			   & $P  \approx 0.04$ & $\Delta F\approx 0.07$ \\ \hline
			  		1-Sym PC & $(P_{\text{eff}}  {\times} \Delta F )_{\text{max}}\approx 5 {\times} 10^{-5}$ 
			  		& $ P_{\text{eff}} \approx 0.0014$ & $ \Delta F \approx 0.03$ \\ \hline \hline
		\end{tabular}
	\end{table}

	\subsection{Teleporting an input squeezed vacuum state}

	We now derive the analytical fidelity expression for 
	teleporting an input squeezed vacuum state with 
	squeezing $\epsilon$ using NGTMSV resource states. The expression of 
	the Wigner characteristic function of squeezed vacuum state is given 
	in  Eq.~(\ref{chi_sqv}) of the Appendix~\ref{intro}. Using the general 
	formula for the fidelity of teleportation~(\ref{fidex1}), we can evaluate 
	the fidelity in this case, which turns out to be
	\begin{equation}\label{ngfids}
		\begin{aligned}
			&F^\text{NG}_\text{sqv}= \bm{\widehat{F}_1} \exp 
			\left(\bm{u}^T M_5 \bm{u} \right),
		\end{aligned}
	\end{equation}
	where the matrix $M_5$ is given in Eq.~(\ref{m5}) of the Appendix~\ref{appex}.
	Setting the limit  $T_1 \rightarrow 1$ and $T_2 \rightarrow 1$ with $m_1=n_1$, $m_2=n_2$    in Eq.~(\ref{ngfids}) yields the 
	fidelity of teleporting an 
	input squeezed vacuum state using TMSV resource state:
	\begin{equation}\label{sqvfidtmsc}
		\begin{aligned}
			F^\text{TMSV}=\left[\left(\frac{1+\tanh{(r+\epsilon)}}{2}\right) \left(\frac{1+\tanh{(r-\epsilon)}}{2} \right)\right]^{1/2}.\\
		\end{aligned}
	\end{equation}
	On putting $\epsilon=0$ in Eq.~(\ref{sqvfidtmsc}), 
	we obtain the fidelity of teleporting an input coherent state using TMSV 
	resource state~(\ref{fidtmsc}).

	\begin{figure}[h!] 
		\begin{center}
			\includegraphics[scale=1]{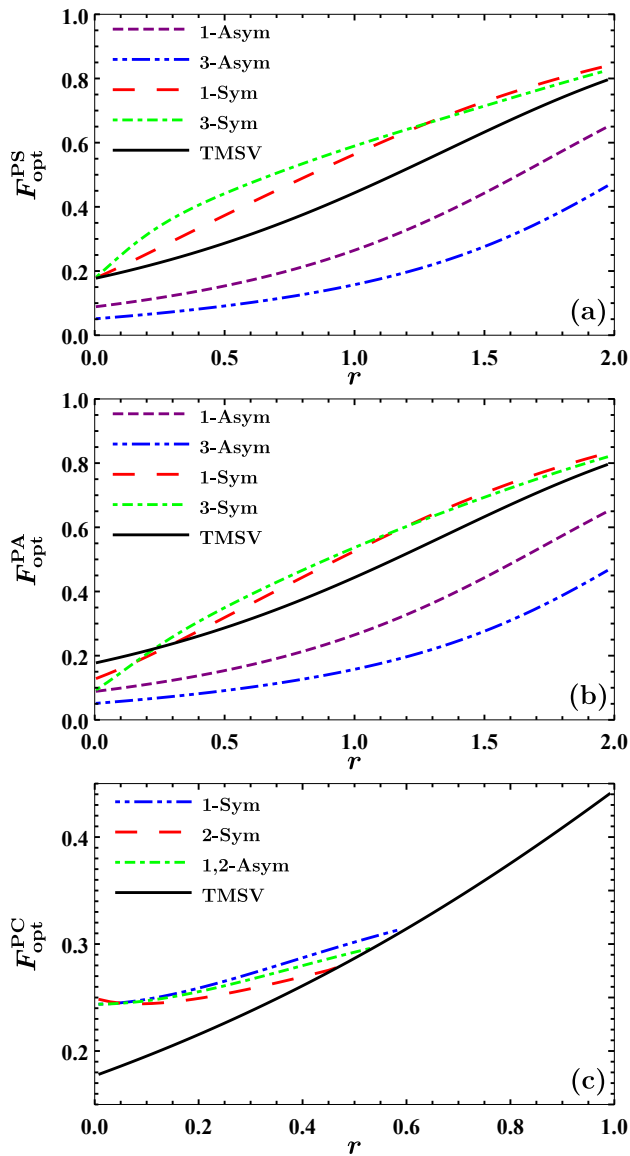}
			\caption{Optimized fidelity for teleporting input squeezed vacuum 
				state as a function 
				of the squeezing parameter for (a) PSTMSV states, 
				(b) PATMSV states, and
				(c) PCTMSV states.  The transmissivities of the
				beam splitters are optimized in order to maximize the fidelity. 
				We have set the squeezing of the input squeezed vacuum state as 
				$\epsilon=1.7$. }
			\label{fids}
		\end{center}
	\end{figure}
	We now numerically analyze the fidelity of teleporting an input squeezed vacuum 
	state. We first optimize the transmissivities of the beam splitters to maximize 
	the fidelity and show the results in Fig.~\ref{fids}. We have taken the squeezing 
	of the input squeezed vacuum state $\epsilon=1.7$. The teleportation results for 
	an input squeezed vacuum state via PSTMSV and PCTMSV resource states show a 
	similar trend as observed for the teleportation of an input coherent state. 
	However, the fidelity of teleportation via symmetric PATMSV resource states 
	outperforms the TMSV state, which contrasts with the results obtained for the 
	teleportation of an input coherent state. The results for the input coherent 
	state are the same as the input squeezed vacuum state with $\epsilon=0$, which 
	represent one extreme, where the PATMSV resource states provide no advantage 
	over the TMSV state. The shown results for $\epsilon=1.7$, which is the 
	maximum achievable squeezing in the lab~\cite{15dB}, represent the other extreme.

\begin{figure}[h!]
		\begin{center}
			\includegraphics[scale=1]{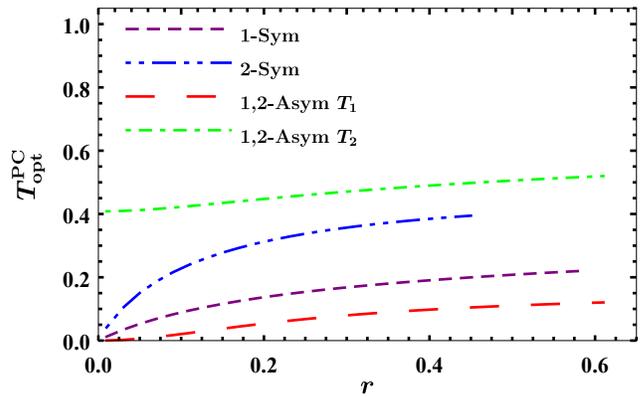}
			\caption{Optimal beam splitter transmissivities as a function 
				of the squeezing parameter for different PCTMSV states.
				The beam splitter transmissivities have been truncated at the 
				minimum squeezing where the TMSV state and the PCTMSV 
				states have the same fidelity. We have set the squeezing of 
				the input squeezed vacuum state as $\epsilon=1.7$.}
			\label{taufids}
		\end{center}
	\end{figure}

	The optimal transmissivity for the PSTMSV and PATMSV  resource states is 
	one; therefore, the results for these cases correspond to ideal PSTMSV and 
	PATMSV resource states, which have been studied in Refs.~\cite{Anno-2007,wang2015}. 
	For different PCTMSV resource states, which have also been discussed in Ref.~\cite{catalysis17}, 
	the corresponding optimal beam splitter transmissivities are shown in 
	Fig.~\ref{taufids}. The behavior of the optimal transmissivities is similar 
	to that of the teleportation of input coherent state [Fig.~\ref{taufidc}].
	
\begin{figure}[h!] 
		\begin{center}
			\includegraphics[scale=1]{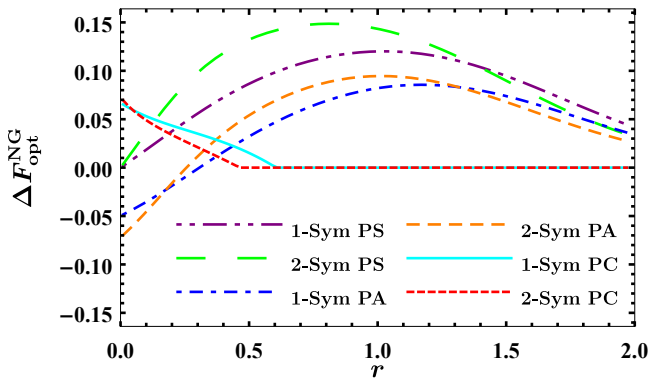}
			\caption{The optimized difference of teleportation fidelity between the
				NGTMSV and the TMSV  states, $\Delta F_{\text{opt}}^{\text{NG}}$,  
				as a function of the squeezing parameter for different non-Gaussian states. 
				The transmissivities have been optimized to maximize $\Delta F^{\text{NG}}$. 
				We have set the squeezing of the input squeezed vacuum state as $\epsilon=1.7$. }
			\label{dfids}
		\end{center}
	\end{figure}

	We now turn to analyze the relative enhancement in performance by considering 
	the difference in teleportation fidelity between
	the NGTMSV states and the TMSV state.
	We show the plots of $\Delta F_{\text{opt}}^{\text{NG}}$ as a function of the 
	squeezing parameter for different NGTMSV states in Fig.~\ref{dfids}. While the
	PSTMSV resource states provide the maximum relative advantage in intermediate 
	squeezing range, the PCTMSV resource states should be preferred in the small 
	squeezing regime.
	
	\subsubsection{Fidelity with squeezing of the input state}\label{sec:epslion}
\begin{figure}[h!] 
		\begin{center}
			\includegraphics[scale=1]{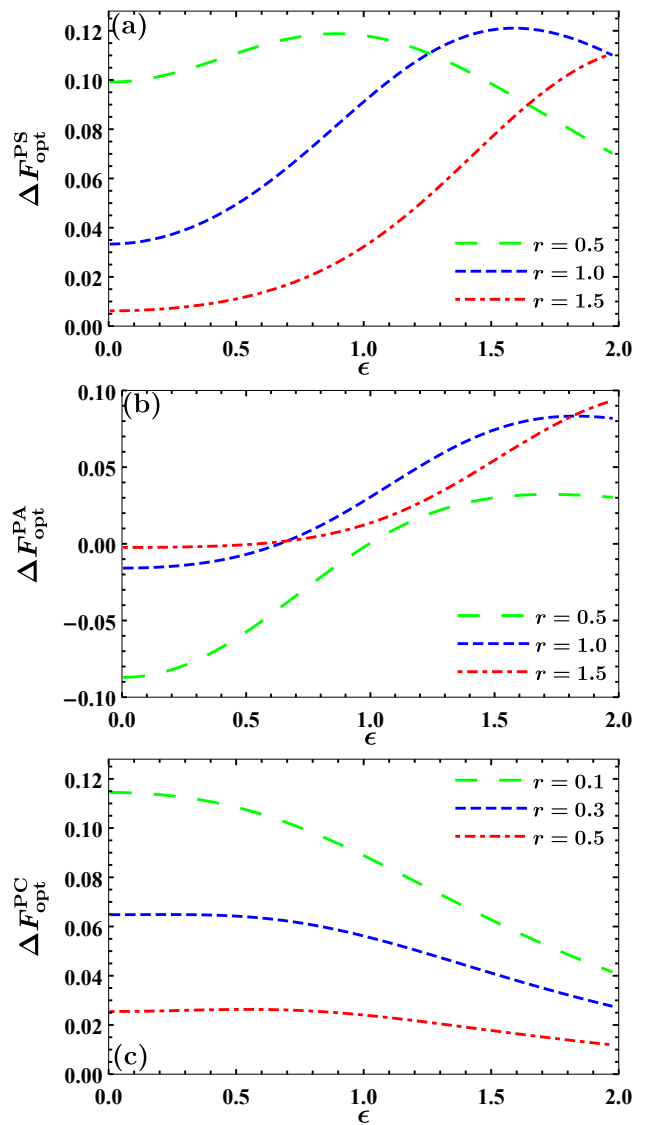}
			\caption{The optimized difference of teleportation fidelity between the
				NGTMSV states and the TMSV  state, $\Delta F_{\text{opt}}^{\text{NG}}$,  as a function 
				of squeezing parameter $\epsilon$ of the input squeezed vacuum state for different squeezing of the TMSV state.  The states considered for different panels are (a) Sym 1-PSTMSV state, (b) Sym 1-PATMSV state, (c) Sym 1-PCTMSV state. The transmissivities have been optimized to maximize the product.}
			\label{rdfids}
		\end{center}
	\end{figure}

	We now analyze the fidelity as a function of the squeezing $\epsilon$ of the input squeezed vacuum state.
	While the absolute fidelity decreases as $\epsilon$ is increased for NGTMSV and TMSV resource states, interesting observations can be made by analyzing $\Delta F^{\text{NG}}_{\text{opt}}$ as a function of $\epsilon$. In Fig.~\ref{rdfids}, we plot $\Delta F_{\text{opt}}^{\text{NG}}$   as a function 
	of squeezing   $\epsilon$ of the input squeezed vacuum state for different squeezing of the resource states. 
	As $\epsilon$ increases in the case of Sym 1-PSTMSV resource state,  the optimized difference $\Delta F^{\text{PS}}_{\text{opt}}$ improves, attains a maximum value, and then starts decreasing.
	The Sym 1-PATMSV  resource state outperforms the TMSV state in a region of $\epsilon$   indicated by the positive value of the optimized difference $\Delta F^{\text{PA}}_{\text{opt}}$. Here too, we notice that as $\epsilon$ increases,  the optimized difference $\Delta F^{\text{PA}}_{\text{opt}}$ improves, attains a maximum value, and then starts decreasing.
	For Sym 1-PCTMSV resource state,  the optimized difference $\Delta F^{\text{PC}}_{\text{opt}}$ continuously decreases as $\epsilon$ is increased.
	One interesting behavior observed for different NGTMSV states is that the fidelity is almost constant for small $\epsilon$, and the inflection in fidelity is observed for higher $\epsilon$. Therefore, the analysis for input coherent state will resemble that of input squeezed vacuum state for small $\epsilon$ ($<0.5$).  
	
	\subsubsection{Relative enhancement in fidelity per trial}
	We now consider the success probability of non-Gaussian operations as well as the probability of the generation of Fock states.
	Since the success probability is independent of $\epsilon$, the analysis in Sec.~\ref{sec:epslion} demonstrates the trend for quantities   $(P {\times} \Delta F)_{\text{opt}}^{\text{NG}}$ or $(P_{\text{eff}} {\times} \Delta F)_{\text{opt}}^{\text{NG}}$ with respect to the squeezing $\epsilon$ of the input squeezed vacuum state.  
	\begin{figure}[h!]
		\begin{center}
			\includegraphics[scale=1]{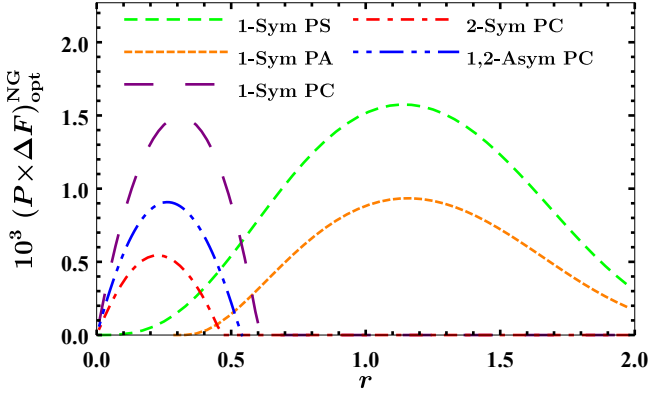}
			\caption{The optimized product $(P {\times} \Delta F)_{\text{opt}}^{\text{NG}}$ as a function of the squeezing parameter. The transmissivities have been optimized to maximize the product. We have set the squeezing of the input squeezed vacuum state as $\epsilon=1.7$. }
			\label{pdfids}
		\end{center}
	\end{figure}

	We now examine $P^{\text{NG}} {\times} \Delta F_{\text{opt}}^{\text{NG}}$ as a function of the squeezing of resource states. The results are shown in Fig.~\ref{pdfids}. 
	We observe that the 1-Sym PS operation provides a maximum advantage when the success probability is taken into account.
	This contrasts with the teleportation of input coherent state, where 1-Sym PC operation provides the maximum advantage.
	However, we note that the 1-Sym PC operation is just a little behind the 1-Sym PS operation in the case of input squeezed vacuum state teleportation.  
	
	\begin{figure}[h!] 
		\begin{center}
			\includegraphics[scale=1]{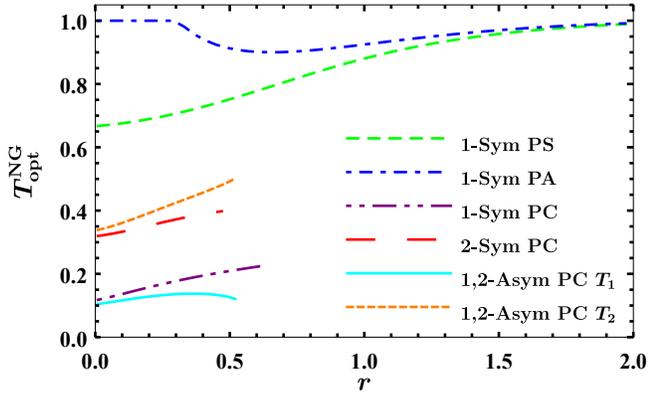}
			\caption{   Optimal beam splitter transmissivities maximizing the product   $P^{\text{NG}} {\times} \Delta F^{\text{NG}}$ as a function
				of the squeezing parameter for different NGTMSV
				states. The beam splitter transmissivities have been
				truncated at the minimum squeezing where the TMSV state and
				the PCTMSV states have the same fidelity. We have set the squeezing of the input squeezed vacuum state as $\epsilon=1.7$.}
			\label{taupdfids}
		\end{center}
	\end{figure}
	
	We have also shown  the optimal beam splitter transmissivities maximizing the product   $P^{\text{NG}} {\times} \Delta F^{\text{NG}}$ as a function of squeezing parameter in Fig.~\ref{taupdfids}.

	\begin{figure}[h!] 
		\begin{center}
			\includegraphics[scale=1]{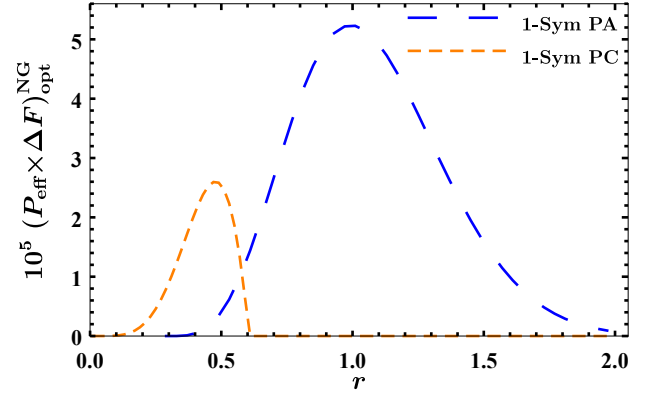}
			\caption{The optimized product    $(P_{\text{eff}} {\times} \Delta F)_{\text{opt}}^{\text{NG}}$ as a function of the squeezing parameter. The transmissivities have been optimized to maximize the product. We have set the squeezing of the input squeezed vacuum state as $\epsilon=1.7$.}
			\label{ppdfids}
		\end{center}
	\end{figure}
	Finally, we plot the optimized product     $(P_{\text{eff}} {\times} \Delta F)_{\text{opt}}^{\text{NG}}$ as a function of the squeezing parameter in Fig.~\ref{ppdfids}. We see that the 1-Sym PA operation outperforms the 1-Sym PC operation.    However, 1-Sym PS operation outperforms both 1-Sym PA and 1-Sym PC operations, which  can be seen explicitly in Table~\ref{table3}.  
	
		\begin{table}[h!]
		\centering
		\caption{\label{table3}
			Maximum value of the product 	$(P {\times} \Delta F)_{\text{opt}}^{\text{NG}}$ and corresponding  $P$ and $\Delta F$. }
		\renewcommand{\arraystretch}{1.5}
		\begin{tabular}{ |c |c |c|c|}
			\hline \hline
			1-Sym PS & $(P {\times} \Delta F )_{\text{max}}\approx 1.6{\times} 10^{-3}$ 
			& $P \approx 0.03$ & $\Delta F \approx 0.05$ \\ \hline
			1-Sym PA & $(P {\times} \Delta F )_{\text{max}}\approx 0.9{\times} 10^{-3}$ 
			& $P \approx 0.03$ & $\Delta F \approx 0.03$ \\ \hline
			1-Sym PA & $(P_{\text{eff}}  {\times} \Delta F )_{\text{max}}\approx 5 {\times} 10^{-5}$ 
			& $ P_{\text{eff}} \approx 0.0016$ & $ \Delta F \approx 0.03$ \\ \hline
			1-Sym PC & $(P {\times} \Delta F )_{\text{max}}\approx 1.5{\times} 10^{-3}$ 
			& $P \approx 0.04$ & $\Delta F \approx 0.03$ \\ \hline
			1-Sym PC & $(P_{\text{eff}}  {\times} \Delta F )_{\text{max}}\approx 2.6 {\times} 10^{-5}$ 
			& $ P_{\text{eff}} \approx 0.0014$ & $ \Delta F \approx 0.02$ \\ \hline \hline
		\end{tabular}
	\end{table}

  While the fidelity is of the
 same order for the considered non-Gaussian operations~\cite{wang2015,catalysis17}, the magnitude of
 $P_{\text{eff}} $ for PA and PC operation is one order less compared to, $P^{\text{NG}}$,  the
 probability of the considered non-Gaussian operations. This renders the PS operation to be more advantageous
 as compared to PA and PC operations.


		\section{Imperfect detectors}\label{sec:imp}
		
		\begin{figure}[H] 
			\begin{center}
				\includegraphics[scale=1]{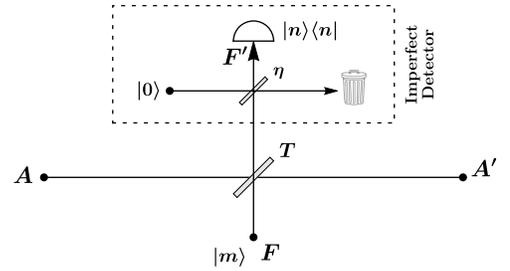}
				\caption{Schematic representation of an imperfect detection in one mode of the TMSV state. An imperfect detector can be modeled via a perfect detector preceded by a beam splitter of transmissivity $\eta$.}
				\label{imperfect}
			\end{center}
		\end{figure}
		In the preceding sections, we have considered the perfect photon detector, \ie, unit quantum efficiency detectors, employed in the implementation of non-Gaussian operations [Fig.~\ref{figsub}]. 
		In this section, we investigate the effects of  imperfect detectors, \ie, non-unit quantum efficiency   detectors, on the teleportation fidelity. An imperfect detector of efficiency $\eta$   (without dark counts)  can be modeled as a perfect detector of unit efficiency preceded by a beam splitter of transmissivity $\eta$~\cite{Braunstein_imperfect,dark}.

\begin{figure}[h!] 
			\begin{center}
				\includegraphics[scale=1]{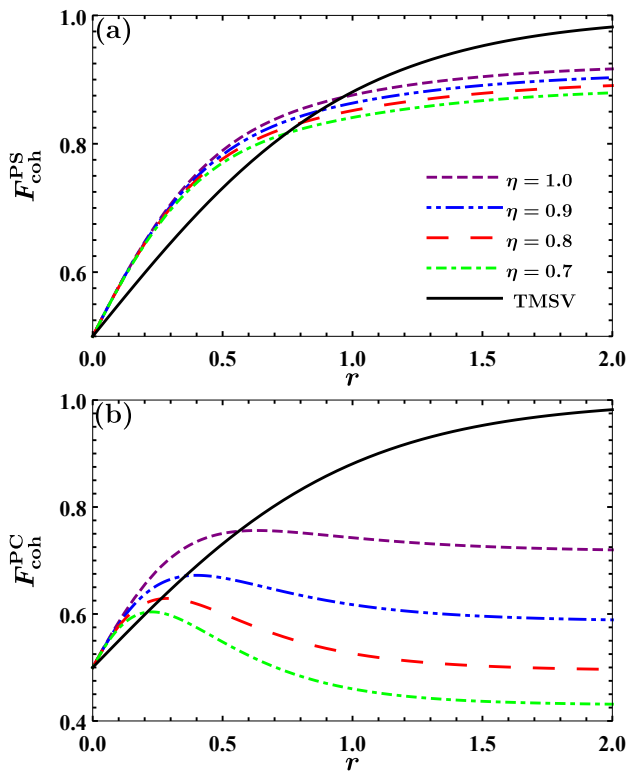}
				\caption{Fidelity of teleporting coherent state as a function of squeezing parameter $r$ for different efficiencies of the detector. The considered resource state is (a) 1-Sym PSTMSV state and (b) 1-Sym PCTMSV state.   The values of the beam splitter transmissivities involved in the non-Gaussian operation have been taken to be (a) $T_1=T_2=0.8$  and (b) $T_1=T_2=0.2$.}
				\label{etafidc}
			\end{center}
		\end{figure}

		We first analyze the fidelity of teleporting coherent state $F^{\text{NG}}_{\text{coh}}$
		as a function of squeezing parameter $r$ for different efficiencies $\eta$
		of the detector in Fig.~\ref{etafidc}. The considered resource states  are  1-Sym
		PSTMSV state and 1-Sym PCTMSV state in  Fig.~\ref{etafidc}(a) and (b), respectively. The results reveal that the fidelity decreases  with decreasing efficiency; however, the fidelity of the non-Gaussian resource states can outperform the TMSV state in a certain squeezing range. 
		We also observe that the detector's inefficiency has a less detrimental effect on the fidelity of 1-Sym PSTMSV than the 1-Sym PCTMSV state. Further, the effect becomes more significant at larger squeezing values.

\begin{figure}[h!] 
			\begin{center}
				\includegraphics[scale=1]{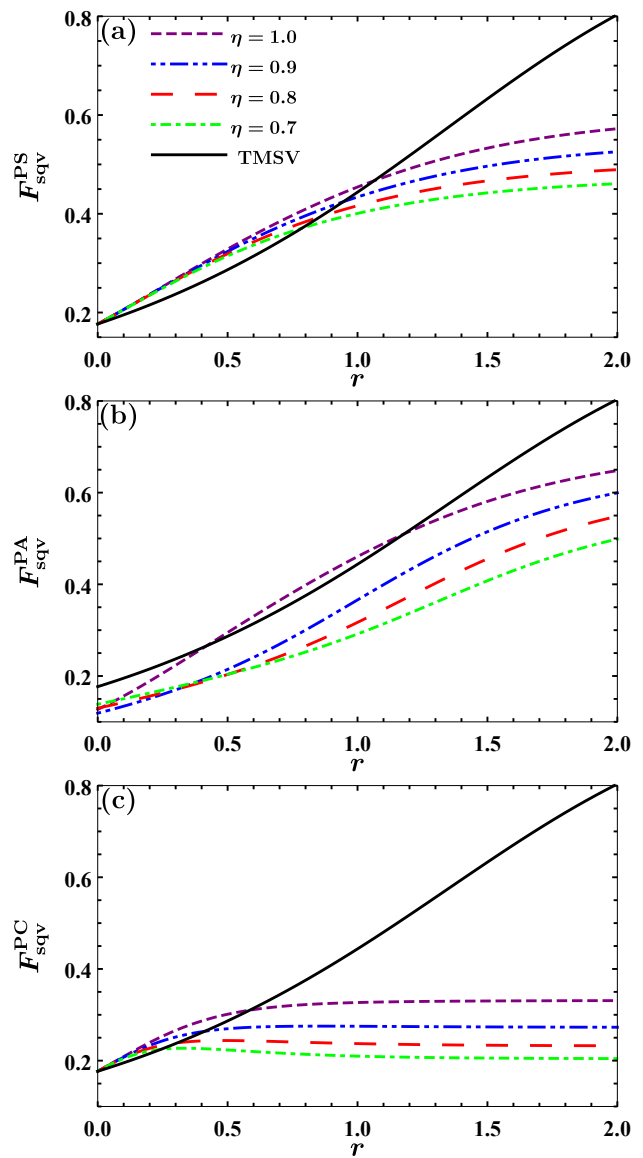}
				\caption{ Fidelity of teleporting   squeezed vacuum state as a function of squeezing parameter $r$ for different efficiencies of the detector. The considered resource state is (a) 1-Sym PSTMSV state, (b) 1-Sym PATMSV state, and (c) 1-Sym PCTMSV state.   The values of the beam splitter transmissivities involved in the non-Gaussian operation have been taken to be (a) $T_1=T_2=0.8$, (b) $T_1=T_2=0.9$  and (c) $T_1=T_2=0.2$.}
				\label{etafids}
			\end{center}
		\end{figure}

		We now turn to the analysis of the fidelity of teleporting squeezed vacuum state $F^{\text{NG}}_{\text{sqv}}$
		as a function of squeezing parameter $r$ for different detector efficiencies $\eta$
		in Fig.~\ref{etafids}.
		The considered resource states are 1-Sym
		PSTMSV state, 1-Sym
		PATMSV state, and 1-Sym PCTMSV state in  Fig.~\ref{etafids}(a), (b), and (c), respectively.
		While 1-Sym PSTMSV   and 1-Sym PCTMSV resource states yield higher fidelity than TMSV states, 1-Sym PATMSV states do not provide any advantages over TMSV even at $\eta=0.9$.

	 If the detector efficiency is low, the teleportation fidelity can drop significantly. However, single photon detectors with over 90\% efficiency are experimentally available~\cite{Lita:08,Marsili2013,Zadeh}, and the corresponding drop in fidelity is not significant; this renders the non-Gaussian operation advantageous.
		In addition to the imperfect photon detectors, imperfect homodyne measurements and dissipation due to interaction with environments need to be taken into account. These problems will be taken up somewhere else.  
	
	
	\section{Conclusion}
	\label{sec:conc}

	We have systematically investigated the optimal conditions for the teleportation of input coherent and squeezed vacuum states using various non-Gaussian resource states while considering the probabilistic preparation of resource states. Our work complements and extends the non-Gaussian quantum teleportation studies performed in Refs.~\cite{Anno-2007,wang2015,catalysis15, catalysis17}.
	While fidelity enhancement using
	NGTMSV resource states was demonstrated in Refs.~\cite{Anno-2007,wang2015,catalysis15, catalysis17}, but the probabilistic considerations were overlooked. Neglecting the success probability can lead to undesirable scenarios.
	For instance, the ideal PSTMSV state (studied in Refs.~\cite{Anno-2007,wang2015}), which maximizes the teleportation fidelity, has a success probability approaching zero (Fig.~\ref{fpcompare}). Therefore, this scenario is highly non-optimal from the point of view of resource utilization.  
	In this article, we have provided a systematic treatment for the optimal performance of the teleportation protocol by trade-off between the teleportation fidelity and the probability of state preparation.

	To summarize the main results, the investigation of the fidelity difference between NGTMSV  and TMSV states shows that the PSTMSV states provide a maximum advantage at intermediate values of squeezing parameters, whereas, at small squeezing, the PCTMSV states provide a maximum advantage.
	Further,   taking the preparation probability of non-Gaussian states into account, we obtain optimal conditions that can be effectively implemented in a lab to enhance fidelity. Furthermore, the symmetric 1-PSTMSV state is the most beneficial resource state among the considered non-Gaussian states. We have also provided the optimal squeezing and beam 
	splitter transmissivity values 
	maximizing the performance. We believe these results will be consequential in the experimental realization of non-Gaussian teleportation protocols.

	Another significant result of our paper is the derivation of a unified 
	analytical expression of the Wigner characteristic function of the NGTMSV states, which does not exist in the literature as far as we know. This derived Wigner characteristic function depends on the transmissivity of the beam splitters and the squeezing of 
	the resource states.
	Further, by choosing the input photon state and the number of detected photons in the auxiliary modes, which also appear in the Wigner characteristic function, one can subtract, add, or catalyze an arbitrary number of photons from the TMSV state. 
	  We also examined the effects of imperfect photon detectors on the teleportation fidelity. While the fidelity decreases due to imperfect detectors, the availability of high efficiency photon detectors renders the  implementation of non-Gaussian operations on TMSV state advantageous.  
	
  The unified Wigner characteristic function will be useful in evaluating  two-mode squeezing~\cite{Hillery}, Mandel-$Q$ parameter~\cite{Mandel:79}, studying antibunching effects~\cite{antibunching}, non-Gaussianity~\cite{non-G} and nonlocality~\cite{nonlocality}   and its considerations 
	in entanglement distillation, entanglement swapping, and quantum illumination 
	protocols. Furthermore, our work may inspire several future investigations identifying optimal conditions in various non-Gaussian QIP protocols.

	\section*{Acknowledgement}
	Both the authors thank Rishabh and Mohak Sharma for a careful reading of the final 
	version of the draft. C.K. acknowledges the financial support from 
	{\bf DST/ICPS/QuST/Theme-1/2019/General} Project number {\sf Q-68}.

	
	\appendix
	
	\section{Brief description of CV systems  and its phase 
		space description}\label{intro}
	Our system of interest is an $n$-mode CV system, whose $i^\text{th}$ mode can be expressed by a pair of Hermitian quadrature operators
	$\hat{q}_i$ and $\hat{p}_i$. We arrange these $n$ pairs in the form of a column vector 
	as~\cite{arvind1995,Braunstein,adesso-2007,weedbrook-rmp-2012,adesso-2014}
	\begin{equation}\label{eq:columreal}
		\hat{ \xi}  =(\hat{ \xi}_i)= (\hat{q_{1}},\,
		\hat{p_{1}}, \dots, \hat{q_{n}}, 
		\, \hat{p_{n}})^{T}, \quad i = 1,2, \dots ,2n.
	\end{equation}
	This permits us to write the bosonic commutation relation  between them   compactly as ($\hbar$=1)
	\begin{equation}\label{eq:ccr}
		[\hat{\xi}_i, \hat{\xi}_j] = i \Omega_{ij}, \quad (i,j=1,2,...,2n),
	\end{equation}
	where $\Omega$ is the 2$n$ $\times$ 2$n$ matrix given by
	\begin{equation}
		\Omega = \bigoplus_{k=1}^{n}\omega =  \begin{pmatrix}
			\omega & & \\
			& \ddots& \\
			& & \omega
		\end{pmatrix}, \quad \omega = \begin{pmatrix}
			0& 1\\
			-1&0 
		\end{pmatrix}.
	\end{equation}
	The photon annihilation and creation 
	operators $\hat{a}_i\, \text{and}\, 
	{\hat{a}_i}^{\dagger}$ ($i=1,2,\, \dots\, ,n$) are given as
	\begin{equation}\label{realtocom}
		\hat{a}_i=   \frac{1}{\sqrt{2}}(\hat{q}_i+i\hat{p}_i),
		\quad  \hat{a}^{\dagger}_i= \frac{1}{\sqrt{2}}(\hat{q}_i-i\hat{p}_i).
	\end{equation}

	We shall be concerned with two symplectic operations discussed below~\cite{arvind1995,weedbrook-rmp-2012}.

	\par
	\noindent{\bf Beam splitter operation:}
	The two mode beam splitter operation acts 
	on the quadrature operators 
	$  \hat{\xi} = (\hat{q}_{i}, \,\hat{p}_{i},\, \hat{q}_{j},\,
	\hat{p}_{j})^{T}$ of a two mode system 
	as follows:
	\begin{equation}\label{beamsplitter}
		B_{ij}(T) = \begin{pmatrix}
			\sqrt{T} \,\mathbb{1}_2& \sqrt{1-T} \,\mathbb{1}_2 \\
			-\sqrt{1-T} \,\mathbb{1}_2& \sqrt{T} \,\mathbb{1}_2
		\end{pmatrix},
	\end{equation}
	where $\mathbb{1}_2$ is the $2 \times 2$ identity matrix. 
	
	\par
	\noindent{\bf  Two mode squeezing operation:}
	The two mode squeezing operation acts  on the quadrature operators 
	$(\hat{q}_{i}$, $\hat{p}_{i}$, $\hat{q}_{j}$, $\hat{p}_{j})^T$ as follows:
	\begin{equation}\label{eq:tms}
		S_{ij}(r) = \begin{pmatrix}
			\cosh r \,\mathbb{1}_2& \sinh r \,\mathbb{Z} \\
			\sinh r \,\mathbb{Z}& \cosh r \,\mathbb{1}_2
		\end{pmatrix},
	\end{equation}
	where $\mathbb{Z} = \text{diag}(1,\, -1)$.
	The TMSV state is obtained by the action of two mode squeezing operator on two single mode vacuum state.

	\subsection{Phase space description}
	Working with Wigner characteristic function turns out to be convenient in quantum teleportation. We can find the Wigner 
	characteristic function corresponding to a density operator 
	$\hat{\rho}$ of an $n$-mode quantum system using the relation
	\begin{equation}\label{wigdef}
		\chi(\Lambda) = \text{Tr}[\hat{\rho} \, \exp(-i \Lambda^T \Omega \hat{\xi})],
	\end{equation}
	where $\xi = (\hat{q_1}, \hat{p_1},\dots \hat{q_n}, \hat{p_n})^T$,  
	$\Lambda = (\Lambda_1, \Lambda_2, \dots \Lambda_n)^T$ with  
	$\Lambda_i = (\tau_i, \sigma_i)^T \in \mathcal{R}^2$.
	For instance, Eq.~(\ref{wigdef}) can be used to evaluate the Wigner characteristic function of a single mode Fock state 
	$|n\rangle$:
	\begin{equation}\label{charfock}
		\chi_{|n\rangle}(\tau,\sigma)=\exp  \left[- \frac{\tau^2}{4}-\frac{\sigma^2}{4} \right]\,L_{n}\left( \frac{\tau^2}{2}+\frac{\sigma^2}{2} \right),
	\end{equation}
	where $L_n(x)$ is the Laguerre polynomial.
	Writing the above equation in terms of exponential generating function, we get 
	\begin{equation}\label{charfock1}
		\chi_{|n\rangle}(\tau,\sigma)=\exp  \left[- \frac{\tau^2}{4}-\frac{\sigma^2}{4} \right]\,	\bm{\widehat{F}}e^{ 2 st +s(\tau+i\sigma)-t(\tau-i\sigma)},
	\end{equation}
	with
	\begin{equation}
		\bm{\widehat{F}} =  \frac{1}{2^n n!}  \frac{\partial^n}{\partial\,s^n} \frac{\partial^n}{\partial\,t^n} \{ \bullet \}_{s=t=0}.
	\end{equation}

	We define the first order moments for an $n$-mode CV system as 
	\begin{equation}
		\bm{d} = \langle  \hat{\xi } \rangle =
		\text{Tr}[\hat{\rho} \hat{\xi}].
	\end{equation}
	Further, the second order moments can be written in the form of a 
	real symmetric $2n\times2n$ matrix, known as covariance matrix:
	\begin{equation}\label{eq:cov}
		V = (V_{ij})=\frac{1}{2}\langle \{\Delta \hat{\xi}_i,\Delta
		\hat{\xi}_j\} \rangle,
	\end{equation}
	where $\Delta \hat{\xi}_i = \hat{\xi}_i-\langle \hat{\xi}_i
	\rangle$, and $\{\,, \, \}$ denotes anti-commutator.

	A special class of states, whose Wigner characteristic function is a  Gaussian, are known as Gaussian states.
	Such states can be uniquely specified via its first and second order
	moments.
	The general formula for the	Wigner characteristic function~(\ref{wigdef}) simplifies as follows for Gaussian states~\cite{weedbrook-rmp-2012, olivares-2012}:
	\begin{equation}\label{wigc}
		\chi(\Lambda) =\exp[-\frac{1}{2}\Lambda^T (\Omega V \Omega^T) \Lambda- i (\Omega \bm{d} )^T\Lambda],
	\end{equation}
	where $\bm{d}$ and $V$ represents the displacement vector and the covariance matrix of the
	Gaussian state. The Wigner characteristic 
	function of a single mode coherent state with displacement 
	$\bm{d}=(d_x,d_p)^T$  evaluates to
	\begin{equation}\label{chi_coh}
		\chi_\text{coh}(\Lambda)= \exp \left[-\frac{1}{4}(\tau ^2+\sigma ^2)-i (\tau  d_p-\sigma  d_x)\right].
	\end{equation}
	The Wigner characteristic function of a single mode squeezed 
	vacuum state turns out to be
	\begin{equation}\label{chi_sqv}
		\chi_\text{sqv}(\Lambda)=\exp \left[-\frac{1}{4}(\tau ^2 e^{2r}+\sigma ^2 e^{-2r}) \right].
	\end{equation}

	Let $\mathcal{U}(S)$ represent the infinite 
	dimensional unitary representation for a homogeneous symplectic transformation $S$. Given the density 
	operator transformation rule as 
	$\rho \rightarrow \,\mathcal{U}(S) \rho
	\,\mathcal{U}(S)^{\dagger}$, the
	transformation of the displacement vector, 
	covariance matrix  and Wigner characteristic function turns out to be~\cite{arvind1995,olivares-2012,weedbrook-rmp-2012}
	\begin{equation}\label{transformation} 
		\bm{d}\rightarrow S \bm{d},\quad V\rightarrow SVS^T,\quad  \text{and} \,\,\chi(\Lambda) \rightarrow \chi(S^{-1}\Lambda).
	\end{equation}

	\section{Matrices appearing in the Wigner characteristic function, and the fidelity of teleportation using NGTMSV states.}\label{appex}
	
	\subsection{Wigner characteristic function of the NGTMSV states}
	Here we provide the explicit forms of the matrices $M_1$, $M_2$ and $M_3$, which appear in the Wigner characteristic function~(\ref{eqchar}) of the NGTMSV states. The matrix $M_1$ is given as
	
	\begin{equation}\label{m1}
		M_1=\frac{-1}{4 a_0}\left(
		\begin{array}{cccc}
			a_1 & 0 & -a_2 & 0 \\
			0 & a_1 & 0 & a_2 \\
			-a_2 & 0 & a_1 & 0 \\
			0 & a_2 & 0 & a_1 \\
		\end{array}
		\right),
	\end{equation}
	where, 
	\begin{equation}
		\begin{aligned}
			a_0=&\beta ^2-\alpha ^2 t_1^2 t_2^2,\\
			a_1=&\beta ^2+\alpha ^2 t_1^2 t_2^2,\\
			a_2=&2 \alpha  \beta  t_1 t_2.\\
		\end{aligned}
	\end{equation}
	Here  $t_i=\sqrt{T_i}$ ($i=1,2$). 
	Further $\alpha=\sinh \, r$ and $\beta=\cosh \, r$. The matrix $M_2$ is given by
	\begin{equation}\label{m2}
		M_2=\frac{1}{a_0}\left(
		\begin{array}{cccc}
			b_1 & i b_1 & b_2 & -i b_2 \\
			-b_1 & i b_1 & -b_2 & -i b_2 \\
			b_3 & -i b_3 & b_4 & i b_4 \\
			-b_3 & -i b_3 & -b_4 & i b_4 \\
			b_5 & i b_5 & b_6 & -i b_6 \\
			-b_5 & i b_5 & -b_6 & -i b_6 \\
			b_7 & -i b_7 & b_8 & i b_8 \\
			-b_7 & -i b_7 & -b_8 & i b_8 \\
		\end{array}
		\right),
	\end{equation}
	where, 
	\begin{equation}
		\begin{array}{ccccccc}
			b_1 & = & \beta ^2 r_1, & \text{   } & b_5 & = & -\alpha ^2 r_1 t_1 t_2^2, \\
			b_2 & = & -\alpha  \beta  r_1 t_1 t_2, & \text{   } & b_6 & = & \alpha  \beta  r_1 t_2, \\
			b_3 & = & -\alpha  \beta  r_2 t_1 t_2, & \text{   } & b_7 & = & \alpha  \beta  r_2 t_1, \\
			b_4 & = & \beta ^2 r_2, & \text{   } & b_8 & = & -\alpha ^2 r_2 t_1^2 t_2. \\
		\end{array}
	\end{equation}
	Here $r_i=\sqrt{1-T_i}$ ($i=1,2$). The matrix $M_3$ is given by
	\begin{equation}\label{m3}
		M_3=\frac{1}{a_0}\left(
		\begin{array}{cccccccc}
			0 & c_1 & c_2 & 0 & 0 & c_3 & c_4 & 0 \\
			c_1 & 0 & 0 & c_2 & c_3 & 0 & 0 & c_4 \\
			c_2 & 0 & 0 & c_5 & c_6 & 0 & 0 & c_7 \\
			0 & c_2 & c_5 & 0 & 0 & c_6 & c_7 & 0 \\
			0 & c_3 & c_6 & 0 & 0 & c_8 & c_9 & 0 \\
			c_3 & 0 & 0 & c_6 & c_8 & 0 & 0 & c_9 \\
			c_4 & 0 & 0 & c_7 & c_9 & 0 & 0 & c_{10} \\
			0 & c_4 & c_7 & 0 & 0 & c_9 & c_{10} & 0 \\
		\end{array}
		\right),
	\end{equation}
	where,
	\begin{equation}
		\begin{array}{ccccccc}
			c_1 & = & \beta ^2 r_1^2, & \text{   } & c_6 & = & -\alpha  \beta  r_1 r_2 t_2, \\
			c_2 & = & \alpha  \beta  r_1 r_2 t_1 t_2, & \text{   } & c_7 & = & \beta ^2 t_2-\alpha ^2 t_1^2 t_2, \\
			c_3 & = & \beta ^2 t_1-\alpha ^2 t_1 t_2^2, & \text{   } & c_8 & = & \alpha ^2 r_1^2 t_2^2, \\
			c_4 & = & -\alpha  \beta  r_1 r_2 t_1, & \text{   } & c_9 & = & \alpha  \beta  r_1 r_2, \\
			c_5 & = & \beta ^2 r_2^2, & \text{   } & c_{10} & = & \alpha ^2 r_2^2 t_1^2. \\
		\end{array}
	\end{equation}
	\subsection{Fidelity for input coherent state using NGTMSV states}
	The explicit form of matrix $M_4$ appearing in the fidelity of teleportation of input coherent state using NGTMSV states~(\ref{ngfidc}) is
	\begin{equation}\label{m4}
		M_4=\frac{1}{d_0}\left(
		\begin{array}{cccccccc}
			0 & c_1 & d_1 & 0 & 0 & d_2 & c_4 & 0 \\
			c_1 & 0 & 0 & d_1 & d_2 & 0 & 0 & c_4 \\
			d_1 & 0 & 0 & c_5 & c_6 & 0 & 0 & d_3 \\
			0 & d_1 & c_5 & 0 & 0 & c_6 & d_3 & 0 \\
			0 & d_2 & c_6 & 0 & 0 & c_8 & d_4 & 0 \\
			d_2 & 0 & 0 & c_6 & c_8 & 0 & 0 & d_4 \\
			c_4 & 0 & 0 & d_3 & d_4 & 0 & 0 & c_{10} \\
			0 & c_4 & d_3 & 0 & 0 & d_4 & c_{10} & 0 \\
		\end{array}
		\right),
	\end{equation}
	where,
	\begin{equation}
		\begin{array}{ccc}
			d_0 & = & 2 \beta \left(\beta -\alpha t_1 t_2\right), \\
			d_1 & = & \beta ^2 r_1 r_2, \\
			d_2 & = & \beta( 2 \beta  t_1-\alpha t_2 \left(t_1^2+1\right)), \\
			d_3 & = & \beta( 2 \beta  t_2-\alpha t_1 \left(t_2^2+1\right)), \\
			d_4 & = & \alpha  r_1 r_2 \left(2 \beta -\alpha  t_1 t_2\right). \\
		\end{array}
	\end{equation} 
	\subsection{Fidelity for input squeezed vacuum state using NGTMSV states}
	The explicit form of matrix $M_5$ appearing in the fidelity of teleportation of input coherent state using NGTMSV states~(\ref{ngfids}) is

	\begin{equation}\label{m5}
		M_5= \frac{1}{e_0} \left(
		\begin{array}{cccccccc}
			e_1 & e_2 & e_3 & e_4 & e_5 & e_6 & e_7 & e_8 \\
			e_2 & e_1 & e_4 & e_3 & e_6 & e_5 & e_8 & e_7 \\
			e_3 & e_4 & e_9 & e_{10} & e_{11} & e_{12} & e_{13} & e_{14} \\
			e_4 & e_3 & e_{10} & e_9 & e_{12} & e_{11} & e_{14} & e_{13} \\
			e_5 & e_6 & e_{11} & e_{12} & e_{15} & e_{16} & e_{17} & e_{18} \\
			e_6 & e_5 & e_{12} & e_{11} & e_{16} & e_{15} & e_{18} & e_{17} \\
			e_7 & e_8 & e_{13} & e_{14} & e_{17} & e_{18} & e_{19} & e_{20} \\
			e_8 & e_7 & e_{14} & e_{13} & e_{18} & e_{17} & e_{20} & e_{19} \\
		\end{array}
		\right),
	\end{equation}
	
	where,
	\begin{equation}
		\begin{aligned}
			e_0=&2 \left(a_0 \delta +a_1\right) ,\\
			e_1=&-\beta ^2 \gamma  r_1^2 ,\\
			e_2=&\beta ^2 r_1^2 \left(\frac{a_0}{a_3}+\delta \right) ,\\
			e_3=&\beta  r_1 r_2 \left(\delta  \left(\frac{a_0}{\sqrt{a_3}}+\alpha  t_1 t_2\right)+\frac{a_1}{\sqrt{a_3}}-\alpha  t_1 t_2\right) ,\\
			e_4=&\beta ^2 \gamma  r_1 r_2 ,\\
			e_5=&-\alpha  \beta  \gamma  r_1^2 t_2 ,\\
			e_6=&\frac{e_0}{a_0} \left(1+\alpha ^2 r_2^2-\alpha  \beta  r_1^2 t_2 \frac{a_0 \delta +a_3}{e_0}\right) ,\\
			e_7=&-\alpha  \beta  r_1 r_2 t_1 \left(\frac{a_0}{a_3}+\delta \right) ,\\
			e_8=&\alpha  \beta  \gamma  r_1 r_2 t_1 ,\\
			e_9=&-\beta ^2 \gamma  r_2^2 ,\\
			e_{10}=&\beta ^2 r_2^2 \left(\frac{a_0}{a_3}+\delta \right) ,\\	
			e_{11}=&-\alpha  \beta  r_1 r_2 t_2 \left(\frac{a_0}{a_3}+\delta \right) ,\\
			e_{12}=&\alpha  \beta  \gamma  r_1 r_2 t_2 ,\\
			e_{13}=&-\alpha  \beta  \gamma  r_2^2 t_1 ,\\
			e_{14}=&\frac{e_0}{a_0} \left(1+\alpha ^2 r_1^2-\alpha  \beta  r_2^2 t_1 \frac{a_0 \delta +a_3}{e_0}\right) ,\\
			e_{15}=&-\alpha ^2 \gamma  r_1^2 t_2^2 ,\\
			e_{16}=&\alpha ^2 r_1^2 t_2^2 \left(\frac{a_0}{a_3}+\delta \right) ,\\
			e_{17}=&\alpha  r_1 r_2 \left(\delta  \left(\frac{2 \beta ^2}{\sqrt{a_3}}-\alpha t_1 t_2 \frac{a_0}{a_3} \right) + \frac{a_1}{\sqrt{a_3}}+\beta \right) ,\\
			e_{18}=&\alpha ^2 \gamma  r_1 r_2 t_1 t_2 ,\\
			e_{19}=&-\alpha ^2 \gamma  r_2^2 t_1^2 ,\\
			e_{20}=&\alpha ^2 r_2^2 t_1^2 \left(\frac{a_0}{a_3}+\delta \right),\\	
		\end{aligned}
	\end{equation}
	with $\gamma=\sinh (2\epsilon)$, $\delta=\cosh (2\epsilon)$ and $a_3=(a_1-a_2)$.

	
	
	
	%

	
\end{document}